# Formation of field induced absorption in the probe response signal of a four-level 'V' type atomic system: a theoretical study

Khairul Islam[1], Arindam Ghosh[1], Dipankar Bhattacharyya[2] and Amitava Bandyopadhyay[1*]

[1]Department of Physics, Visva-Bharati, Santiniketan, PIN 731 235, West Bengal, India

[2]Department of Physics, Santipur College, Santipur, Nadia, PIN 741 404, West Bengal, India

**Abstract:** A density matrix based analytical model is developed to study the coherent probe field propagation through a four-level 'V' type system in presence of a coherent control field. The model allows coupling of the probe field from the upper ground level to both of the excited levels keeping the control field locked to a particular transition. The addition of an extra ground level to a conventional three-level 'V' type system creates extra decay paths to the ground levels for the upper level population. A set of sixteen density matrix based equations are formed and then solved analytically under rotating wave approximation to study the probe response under steady state condition. The simulated probe absorption spectra shows absorption dip at the centre of a transparency window only under Doppler broadened condition although the conventional EIT window appears under Doppler free condition. The dependence of the field induced absorption signal on the Rabi frequency of the control field, population transfer rate among the ground levels and temperature of the vapour medium has been studied in details.

*Corresponding author

Email: khislam1986@gmail.com (K. Islam), arindam.ghosh.phy@gmail.com (A. Ghosh), bh.dipankar@gmail.com (D. Bhattacharyya), m2amitava@gmail.com (A. Bandyopadhyay).

## 1. Introduction

Electromagnetically induced transparency (EIT) [1-5], a prime example of quantum interference effect, has been the centre of attraction to a large section of researchers who are engaged in the field of quantum optics. When the frequency difference between the control and the probe beams is set equal to the separation between the ground levels or two excited levels of a multi-level system under consideration, EIT occurs [2] and an otherwise absorptive medium becomes transparent to a resonant (or detuned) low power coherent probe beam in presence of a coherent high-power control beam due to the formation of 'dark state'. Alternatively, depending upon the choice of level schemes, this may also be explained on the basis of the splitting of the energy level, which is common to both the probe and the pump beams, as seen by the probe beam [2]. The reason behind the worldwide attraction of EIT among researchers is of course the potential application of EIT in developing future optical devices that are supposed to be used in optical logic gates, all optical switches [6,7], optical delay generators [8] etc. In addition to this, pure academic interest is also driving researchers to investigate this effect using various level schemes like inverted-Y [9, 10], four-level cascade [10], M and N-type systems [11, 12] etc. The effect of thermal velocity averaging also influences the EIT line shape in a way so as to reduce its width compared to the Doppler free condition. The Doppler broadening affects the dispersive properties of a medium substantially. The extremely low line width of the EIT window under Doppler broadened regime corresponds to a very steep dispersion that results in very low group velocity of a probe pulse passing through a medium in presence of high power control field. This phenomenon points towards a possibility to manipulate the probe pulse propagation through an atomic vapour medium. In fact, the theoretical prediction [13] has been well supported by the experimental demonstration of reduction of the group velocity of a probe pulse propagating through hot $^{87}Rb$ vapour [14]. The deceleration and storage of a light pulse in an atomic vapour and then its release on demand ultimately fulfilled the dream of stopping and storage of light. The observation of EIT is most easily realized in a three-level $\Lambda$ type system [15] although in reality it is almost impossible to form a true three-level $\Lambda$ type system even in the alkali atomic vapour medium. All most all alkali atoms have complex multilevel atomic structure. Many reports in this regard can be found in the literature. Vasant Natarajan et. al. [16] has demonstrated experimentally how the Doppler averaging in a rubidium vapour cell at room temperature reduces the EIT line width to sub-natural value for a $\Lambda$ type system. They have also presented a simple theoretical model in this connection. Very recently D.

Bhattacharyya et. al. [17] showed the formation of EIT with sub-natural line width in a six-level Λ type system in $^{87}$Rb as well as $^{85}$Rb vapour at room temperature. An analytical model has been presented there to explain the sub-natural line width of the observed EIT signal. They have also stated there that the inclusion of all the hyperfine levels in the theoretical modelling has helped them to reproduce the experimentally observed spectra. But in three-level 'V' and cascade type systems too EIT can be produced. D. J. Fulton et. al. [15] reported a comparative study of EIT in 'V', Λ and cascade (Ξ) type systems along with their experimental findings. They have explained why the Λ type system is the most well suited to study EIT on the basis of coherence dephasing rate between the dipole-forbidden transitions. There has been very large number of reports on both the theoretical and experimental studies on EIT [15-17 and the references in 17] as well as electromagnetically induced absorption (EIA) [18, 19] using Λ type level scheme. Although the three-level systems serve well in understanding the physical picture behind formation of EIT, these are hard to be found in reality. Formation of hyperfine levels in atoms almost everywhere makes the level scheme complicated enough to consider more than three energy levels in the theoretical models in order to explain the experimentally observed spectra. In this report, we shall present a theoretical investigation on the propagation of a resonant weak probe field through a four-level 'V' type atomic system in presence of a strong coupling field known as the pump field or control field. The treatment is completely analytic and the simulated probe absorption spectra exhibits formation of a field induced absorption dip on the background of an EIT peak under the Doppler broadened regime whereas under Doppler free condition a transparency window in the probe absorption profile is created. We shall also demonstrate how the different velocity groups of atoms contribute to the probe absorption profile and form the absorption dip. The dispersion property of the system will be studied too.

## 2. Theoretical Model:

We shall now describe the four-level system in details. There are two ground levels and two close upper levels. We have set the separation between the two excited states |3> and |4> to be equal to 266.6 MHz and that between the ground levels as 6.8 GHz following the energy level of $^{87}$Rb [20]. The frequency of the probe beam is scanned from the upper ground state |2> to the excited states |3> and |4>. Hence it couples both the upper states |3> and |4> to the ground level |2>. The frequency of the pump beam is kept locked between the levels |2> and |3>. The coupling of the probe beam to both the upper levels makes the formulation different

from the conventional 'V' type system where the probe field couples one of the upper states from the ground level and the control (pump) field acts between the ground level and the other upper level. The spontaneous decay of the population from the uppermost level |4> to the ground level |2> is dipole-allowed whereas that of population from the uppermost level |4> to the lowest ground level |1> may or may not be dipole-allowed. We assume that the population of the level |3> can always decay spontaneously to both the ground levels |1> and |2>. The provision of population transfer between the two ground levels has been kept in the theoretical formulation so that the collision induced transfer of population among the ground levels may be taken into consideration if so desired. Fig. 1 below shows the level diagram schematically. The population decay rates of the level |*i*> ($i$ = 3, 4) to level | *j*> ($j$ = 1, 2) have been represented by $\gamma_{ij}$ whereas the circular frequencies of probe and control fields are symbolized by $\omega_p$ and $\omega_c$ respectively.

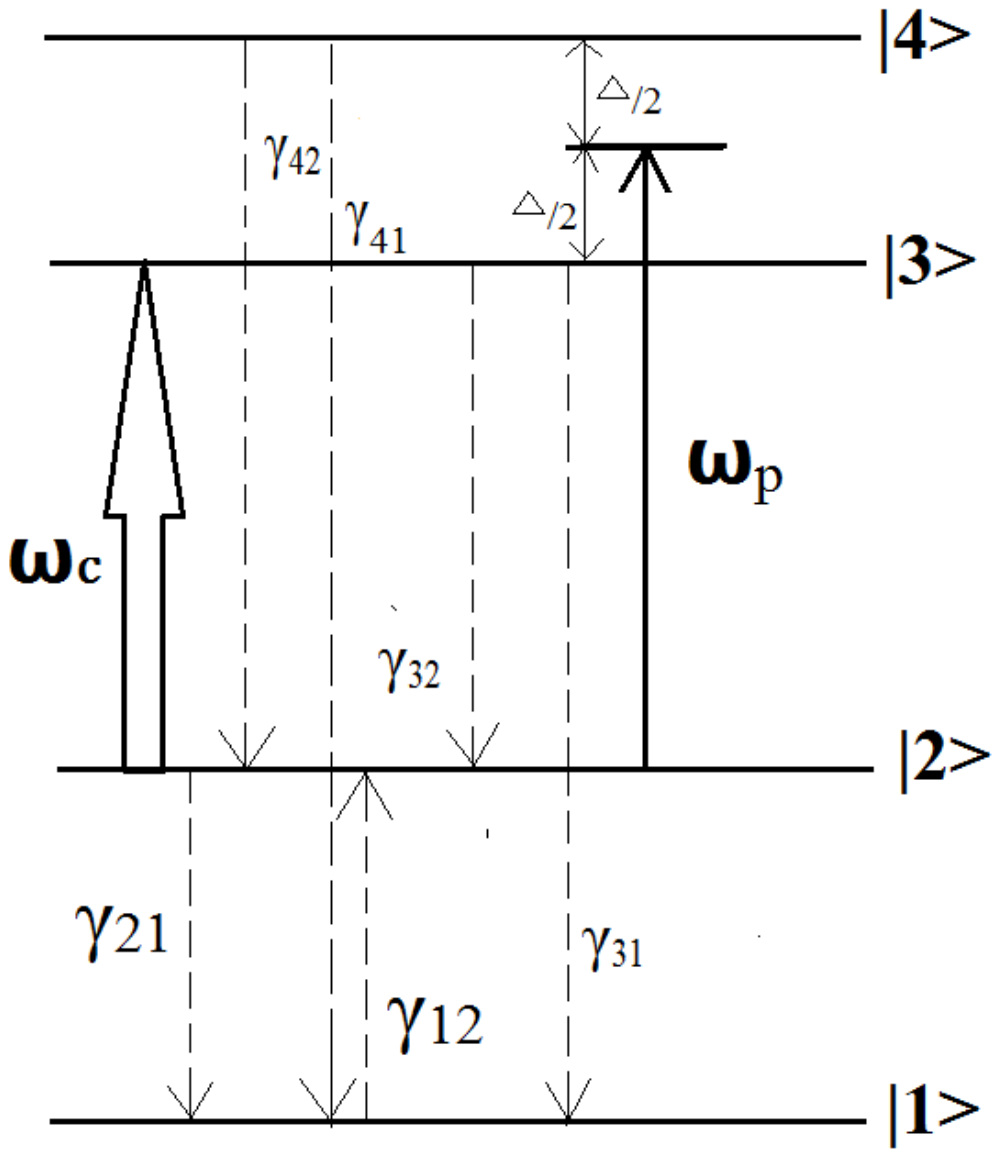

Fig.1: Schematic diagram of the level scheme. The dotted lines represent the decay of population. The frequencies of the control and probe fields are represented by $\omega_c$ and $\omega_p$.

Now we can write the Hamiltonian of the above level scheme in the following manner,

$H = H_0 + H_I$.......................................................................... (1)

Here $H$, $H_0$ and $H_I$ represent the total Hamiltonian, unperturbed Hamiltonian and the interaction Hamiltonian of the system.

$H_0 = \hbar\sum_{i=1}^{4} \omega_i\, |i><i| = \hbar\omega_1|1><1| + \hbar\omega_2|2><2| + \hbar\omega_3|3><3| + \hbar\omega_4|4><4|$. .................. (2)

$H_I = -x_1\frac{\hbar\Omega_c}{2}\{|2><3|e^{i\omega_c t} + |3><2|e^{-i\omega_c t}\} - x_1\frac{\hbar\Omega_p}{2}\{|2><3|e^{i(\omega_p - \Delta/2)t} + |3><2|e^{-i(\omega_p - \Delta/2)t}\}$

$-x_2 \frac{\hbar\Omega_p}{2}\{|2><4|e^{i(\omega_p+\Delta/2)t}+|4><2|e^{-i(\omega_p+\Delta/2)t}\}$..........................................(3)

Here the control and probe Rabi frequencies are defined by $\Omega_c = \mu_{ij}.E_c/\hbar$ and $\Omega_p = \mu_{ij}.E_p/\hbar$. The transition ($|i> \longrightarrow |j>$) dipole matrix element is $\mu_{ij}$. $E_{c(p)}$ is the amplitude of the applied control (probe) field. $\Delta = (\omega_4 - \omega_3) = 266.6$ MHz, where $\hbar\omega_i$ corresponds to the energy of the $i^{th}$level ($i$ = 1, 2, 3, 4).The relative strengths of the two transitions $|2> \rightarrow |3>$ and $|2> \rightarrow |4>$ are given by $x_1$ and $x_2$ respectively. A set of sixteen equations involving the population and coherence terms of the four-level system are derived by using the Liouville's equation of motion [21, 22]

$$\frac{d\rho}{dt} = -\frac{i}{\hbar}[H,\rho] + \Lambda \quad (4)$$

with, $\Lambda = \frac{\Gamma}{2}[\sigma_+\sigma_-\rho + \rho\sigma_-\sigma_+ - 2\sigma_+\rho\sigma_-]$, $\Gamma$ stands for the atomic decay rates. The atomic transition operators are given by$\sigma_+$ and $\sigma_-$and they are complex conjugate of each other [21, 22]. The population and the coherence terms are (diagonal and off-diagonal elements of the density matrix respectively) obtained by solving the following equations and their complex conjugates analytically following rotating wave approximation [21, 22]:

$$\gamma_{21}\rho_{22} + \gamma_{31}\rho_{33} + \gamma_{41}\rho_{44} - \gamma_{12}\rho_{11} = 0 \quad (5)$$

$$x_1\frac{i\Omega_p}{2}\{\rho_{32}^p - \rho_{23}^p\} + x_1\frac{i\Omega_c}{2}\{\rho_{32}^c - \rho_{23}^c\} + x_2\frac{i\Omega_p}{2}\{\rho_{42}^p - \rho_{24}^p\}$$
$$+ \gamma_{12}\rho_{11} - \gamma_{21}\rho_{22} + \gamma_{32}\rho_{33} + \gamma_{32}\rho_{33} = 0 \quad (6)$$

$$x_1\frac{i\Omega_c}{2}\{\rho_{23}^c - \rho_{32}^c\} - (\gamma_{31} + \gamma_{32})\rho_{33} + \gamma_{43}\rho_{44} = 0 \quad (7)$$

$$x_2\frac{i\Omega_p}{2}\{\rho_{24}^p - \rho_{42}^p\} - (\gamma_{41} + \gamma_{42} + \gamma_{43})\rho_{44} = 0 \quad (8)$$

$$(\Gamma_{32}+i\Delta_{32}^c)\rho_{23}^c - x_1\frac{i\Omega_c}{2}(\rho_{33} - \rho_{22}) + (\Gamma_{32}+i\Delta_{32}^p)\rho_{23}^p - x_1\frac{i\Omega_p}{2}(\rho_{33} - \rho_{22}) - x_2\frac{i\Omega_p}{2}\rho_{43} = 0 \quad (9)$$

$$(\Gamma_{42}+i\Delta_{42}^p)\rho_{24}^p - x_2\frac{i\Omega_p}{2}(\rho_{44} - \rho_{22}) - x_1\frac{i\Omega_c}{2}\rho_{34} - x_1\frac{i\Omega_p}{2}\rho_{34} = 0 \quad (10)$$

$$\{\Gamma_{43}+i(\Delta_{42}^p - \Delta_{32}^c)\}\rho_{34} - x_1\left(\frac{i\Omega_p}{2} + \frac{i\Omega_c}{2}\right)\rho_{24}^p + x_2\frac{i\Omega_p}{2}\rho_{32}^c + x_2\frac{i\Omega_p}{2}\rho_{32}^p = 0 \quad (11)$$

We have considered, $\rho_{23}(t) = \rho_{23}^c e^{i\omega_c t} + \rho_{23}^p e^{i(\omega_p-\Delta/2)t}$, $\rho_{24}(t) = \rho_{24}^p e^{i(\omega_p+\Delta/2)t}$ subject to the boundary condition $\rho_{11} + \rho_{22} + \rho_{33} + \rho_{44} = 1$. The probe absorption and dispersion are determined by using the imaginary and real parts of the coherence terms induced by the probe between the levels $|2>$ and $|3>$ ($\rho_{23}^p$), $|2>$ and $|4>$ ($\rho_{24}^p$) [17, 21, 22]. The analytical expression of the $\rho_{23}^p$ and $\rho_{24}^p$ can be derived by solving sixteen optical Bloch equations (OBE) under steady state condition.

$$\rho_{23}^p = \frac{1}{a_3^2+a_4^2}[-x_1x_2^2\frac{\Omega_p^2}{8}(\Omega_p + \Omega_c)(a_4 + ia_3)(\rho_{44} - \rho_{22})$$

$+x_1 \frac{(\Omega_c+\Omega_p)}{2}(a_2 + ia_1)(a_3 - ia_4)(\rho_{33} - \rho_{22}) - (a_5 + ia_6)(a_3 - ia_4)]$ ..........................(12)

$\rho_{24}^{p} = \frac{1}{a_{15}^2+a_{16}^2}[x_2 \frac{\Omega_p}{2}(a_{13} + ia_{14})(a_{16} + \mathrm{i}a_{15})(\rho_{44} - \rho_{22}) - x_1^2 x_2 \frac{\Omega_\mathrm{p}}{2}\left(\Omega_p + \Omega_c\right)^2 (a_{16} + ia_{15})$

$(\rho_{33} - \rho_{22}) + x_1 x_2 \frac{\Omega_p}{2}(\frac{\Omega_c}{2} + \frac{\Omega_p}{2})(\Delta_{32}^{c} - \Delta_{32}^{p})(a_{16} + ia_{15})\{Re(\rho_{23}^{c}) - iIm(\rho_{23}^{c})\}]$....... (13)

with, $Im(\rho_{23}^{c}) = \frac{x_1 \frac{\Omega_c}{2}(\Gamma_{32})}{[\Gamma_{32}^2+(\Delta_{32}^{c})^2]}(\rho_{33}^{0} - \rho_{22}^{0})$, $Re(\rho_{23}^{c}) = \frac{x_1 \frac{\Omega_c}{2}(\Delta_{32}^{c})}{[\Gamma_{32}^2+(\Delta_{32}^{c})^2]}(\rho_{33}^{0} - \rho_{22}^{0})$

$a_1 = \Gamma_{42}\Gamma_{43} - \Delta_{42}^{p}(\Delta_{42}^{p} - \Delta_{32}^{c}) + \frac{x_1^2}{4}\left(\Omega_p + \Omega_c\right)^2$

$a_2 = \Gamma_{42}(\Delta_{42}^{p} - \Delta_{32}^{c}) + \Gamma_{43}\Delta_{42}^{p}$

$a_3 = a_1\Gamma_{32} + a_2\Delta_{32}^{p} + x_2^2 \frac{\Omega_p^2}{4}\Gamma_{42}$, $a_4 = a_1\Delta_{32}^{p} - a_2\Gamma_{32} - x_2^2 \frac{\Omega_p^2}{4}\Delta_{42}^{p}$

$a_5 = a_1\Gamma_{32} + a_2\Delta_{32}^{c} + x_2^2 \frac{\Omega_p^2}{4}\Gamma_{42}$, $a_6 = a_1\Delta_{32}^{c} - a_2\Gamma_{32} - x_2^2 \frac{\Omega_p^2}{4}\Delta_{42}^{p}$

$a_7 = \frac{(\Omega_c+\Omega_p)}{2}\frac{a_2 a_3 + a_1 a_4}{a_3^2+a_4^2}$, $a_8 = \frac{(\Omega_c+\Omega_p)}{2}\frac{a_1 a_3 - a_2 a_4}{a_3^2+a_4^2}$

$a_9 = \frac{x_1 x_2^2 \frac{\Omega_p^2}{8}(\Omega_p+\Omega_c)a_4}{a_3^2+a_4^2}$, $a_{10} = \frac{x_1 x_2^2 \frac{\Omega_p^2}{8}(\Omega_p+\Omega_c)a_3}{a_3^2+a_4^2}$

$$a_{11} = \frac{Re(\rho_{23}^{c})(a_3 a_5 + a_4 a_6) - Im(\rho_{23}^{c})(a_3 a_6 - a_4 a_5)}{a_3^2 + a_4^2}$$

$$a_{12} = \frac{Re(\rho_{23}^{c})(a_3 a_6 - a_4 a_5) + Im(\rho_{23}^{c})(a_3 a_5 + a_4 a_6)}{a_3^2 + a_4^2}$$

$a_{13} = \Gamma_{32}\Gamma_{43} + \Delta_{32}^{p}(\Delta_{42}^{p} - \Delta_{32}^{c}) + \frac{x_2^2}{4}\Omega_p^2$, $a_{14} = \Gamma_{32}(\Delta_{42}^{p} - \Delta_{32}^{c}) - \Gamma_{43}\Delta_{32}^{p}$

$a_{15} = a_{13}\Gamma_{42} - a_{14}\Delta_{42}^{p} + x_1^2 \frac{\left((\Omega_p+\Omega_c)^2\right)}{4}\Gamma_{32}$, $a_{16} = a_{13}\Delta_{42}^{p} + a_{14}\Gamma_{42} - x_1^2 \frac{\left((\Omega_p+\Omega_c)^2\right)}{4}\Delta_{32}^{p}$

The nature of probe absorption and probe dispersion can be obtained by simulating ($\rho_{23}^{p} + \rho_{24}^{p}$) and then plotting its imaginary and real parts separately as functions of probe detuning ($\Delta_{42}^{p}$).

## 3. Simulation:

The analytical expressions for ($\rho_{23}^{p} + \rho_{24}^{p}$) given in Eq.(12) and Eq.(13) have been used to simulate the probe absorption and probe dispersion under different conditions. We have noticed distinctly different absorptive and dispersive properties of the atomic system under Doppler free and Doppler broadened condition. We have used the data available in the literature for the $D_2$ transition of $^{87}$Rb [20] atoms in the simulation process in order to make the theoretical study realistic and useful to the experimentalists as well. The spontaneous decay rates ($\gamma_{ij}$, $i$ = 3, 4; $j$ = 1,2) of population from both the upper levels to the ground levels have been taken to be 6 MHz each although we have put $\gamma_{41}$= 0, i. e. the spontaneous decay of population from |4> to |1> is kept dipole forbidden in most of this work if not specified otherwise. The control field has been kept locked to the transition from |2> to |3>, hence the corresponding detuning of the control field ($\Delta_{32}^{c}$) is kept equal to zero during the entire simulation process. We shall first compare the simulated probe absorption spectra under Doppler free condition with that under Doppler broadened regime. The probe Rabi frequency ($\Omega_p$) has been kept equal to 1 MHz whereas the Rabi frequency ($\Omega_c$) of the control field has been varied from 10 MHz to 50 MHz in steps of 10MHz. The difference in the absorption line shape under the two conditions is visible. At zero probe field detuning ($\Delta_{42}^{p}$ = 0, $\omega_p = \omega_4 - \omega_2$), the control and the probe fields are on resonant to the transitions |2>⟶ |3> and |2> ⟶ |4> respectively. A 'V' type system is formed and EIT signal in the probe absorption is expected to appear around $\Delta_{42}^{p}$ = 0. But in addition to a transparency peak, an absorption dip at the middle of the transparency window is also visible at $\Delta_{42}^{p}$ = 0 under Doppler broadened condition (Fig.2). We shall term this as field induced absorption.

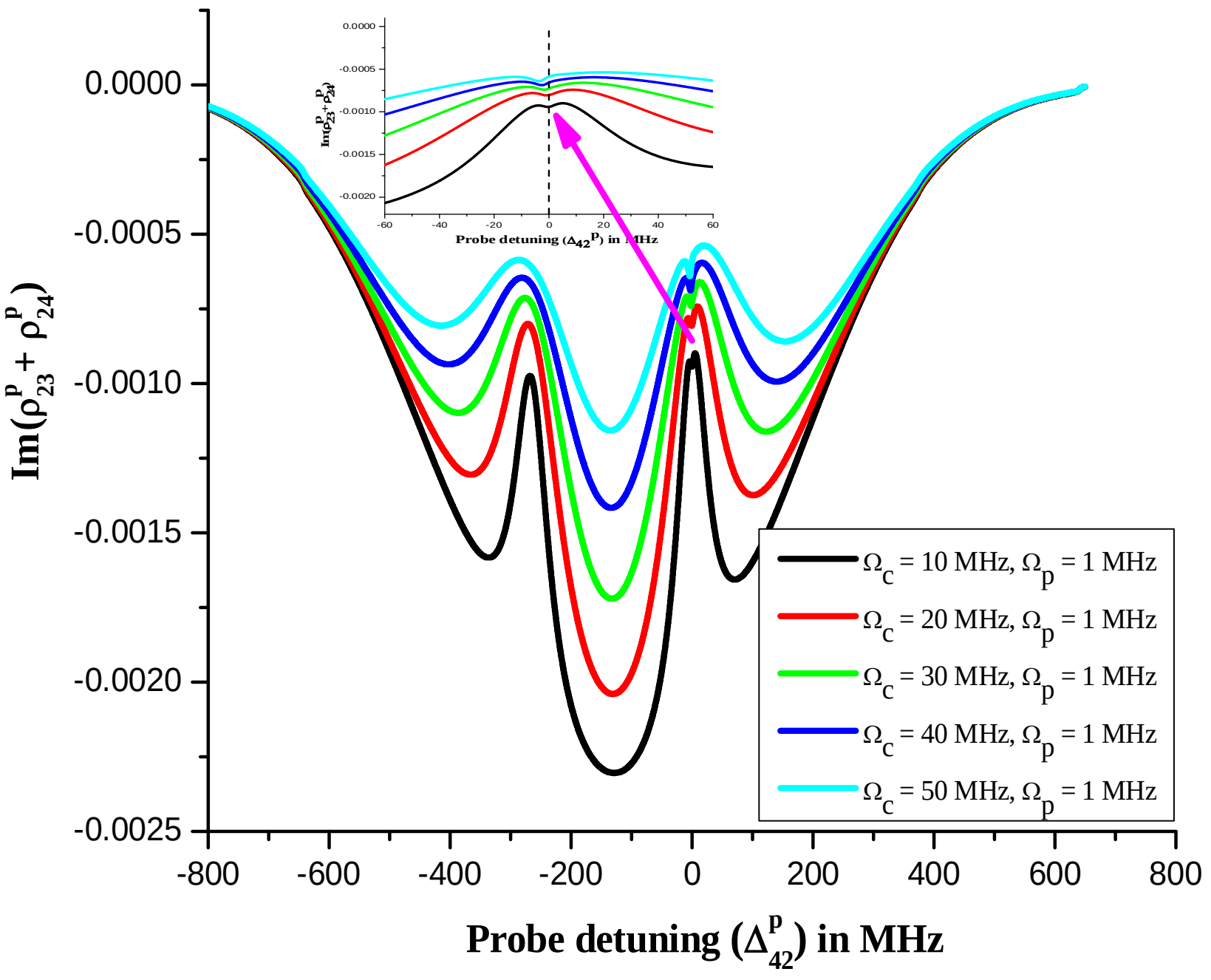

Fig.2: Plot of imaginary part of ($\rho_{23}^{p}$+ $\rho_{24}^{p}$) vs. probe frequency detuning ($\Delta_{42}^{p}$) under the Doppler broadened regime (T = 300 K). The values of the control ($\Omega_c$) and probe ($\Omega_p$) Rabi frequencies have been mentioned in the figure.

The absorption dip created on the background of the transparency window under Doppler broadened condition is found to enhance with increase in the Rabi frequency of the control field. At the same time a regular shift of the absorption dip on the background of the transparency window can also be noticed with increase in the control Rabi frequency (inset of Fig.2). But under Doppler free regime we are only getting two absorption lines in the simulated probe response. The separation between these two absorption dips is equal to the separation between the two upper levels |3> and |4> (266.6 MHz)with a transparency window appearing at $\Delta_{42}^{p}$ = 0on the background of the absorption dip when the frequency difference between the control and probe fields equals the frequency difference between the upper levels |3> and |4> (Fig.3). This is the usual EIT signal.

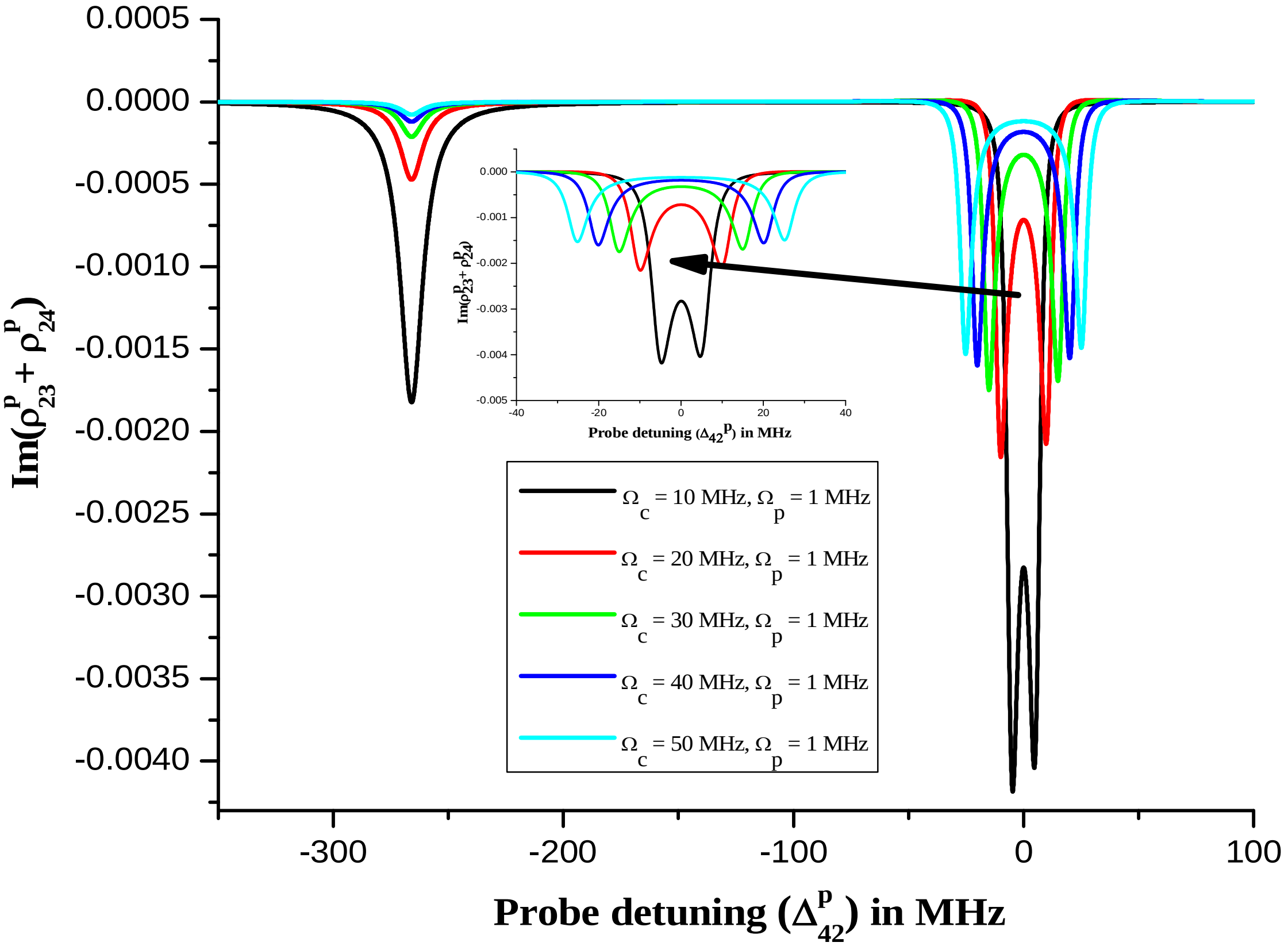

Fig.3: Plot of imaginary part of the ($\rho_{23}^{p}$+ $\rho_{24}^{p}$) vs. probe frequency detuning ($\Delta_{42}^{p}$) under the Doppler free condition. The values of the control ($\Omega_c$) and probe ($\Omega_p$) Rabi frequencies have been mentioned in the figure.

A transparency like peak at $\Delta_{42}^{p} = -266.6$ MHz is generated in the Doppler broadened probe response spectra (Fig.2) but the Doppler free condition shows only an absorption dip at this value of the probe detuning. At this condition, both the pump and the control fields are on-resonant with the |2> → |3> transition. The probe absorption decreases here since the control

field creates a hole in the population of the zero velocity groups of atoms, hence the probe field would see less population to interact with. This is similar to saturation absorption spectroscopy and formations of Lamb dip [23]. The width of this EIT like peak created under Doppler broadened condition is seen to increase with increase in the control Rabi frequency (Fig.2).The formation of an absorption dip on the background of a transparency window in the probe absorption spectra and the corresponding shift with increase in the control Rabi frequency are present only under Doppler broadened regime (please compare Fig. 2 and Fig. 3 around $\Delta_{42}^{p}$ = 0). These features cannot be observed in the probe absorption spectra for a simple three-level 'V' type system [24]. The presence of an extra ground level (|1> in this case) provides an extra decay channel to the population pumped to the upper levels. To get an idea of the contribution of the non-zero velocity groups of atoms towards the probe response we have plotted imaginary part of ($\rho_{23}^{p}$+ $\rho_{24}^{p}$) vs. probe detuning ($\Delta_{42}^{p}$) for specific velocity groups (Fig.4). At zero probe field detuning, the contribution towards the probe absorption by different velocity groups adds up to create the observed absorption dip in Fig.2.

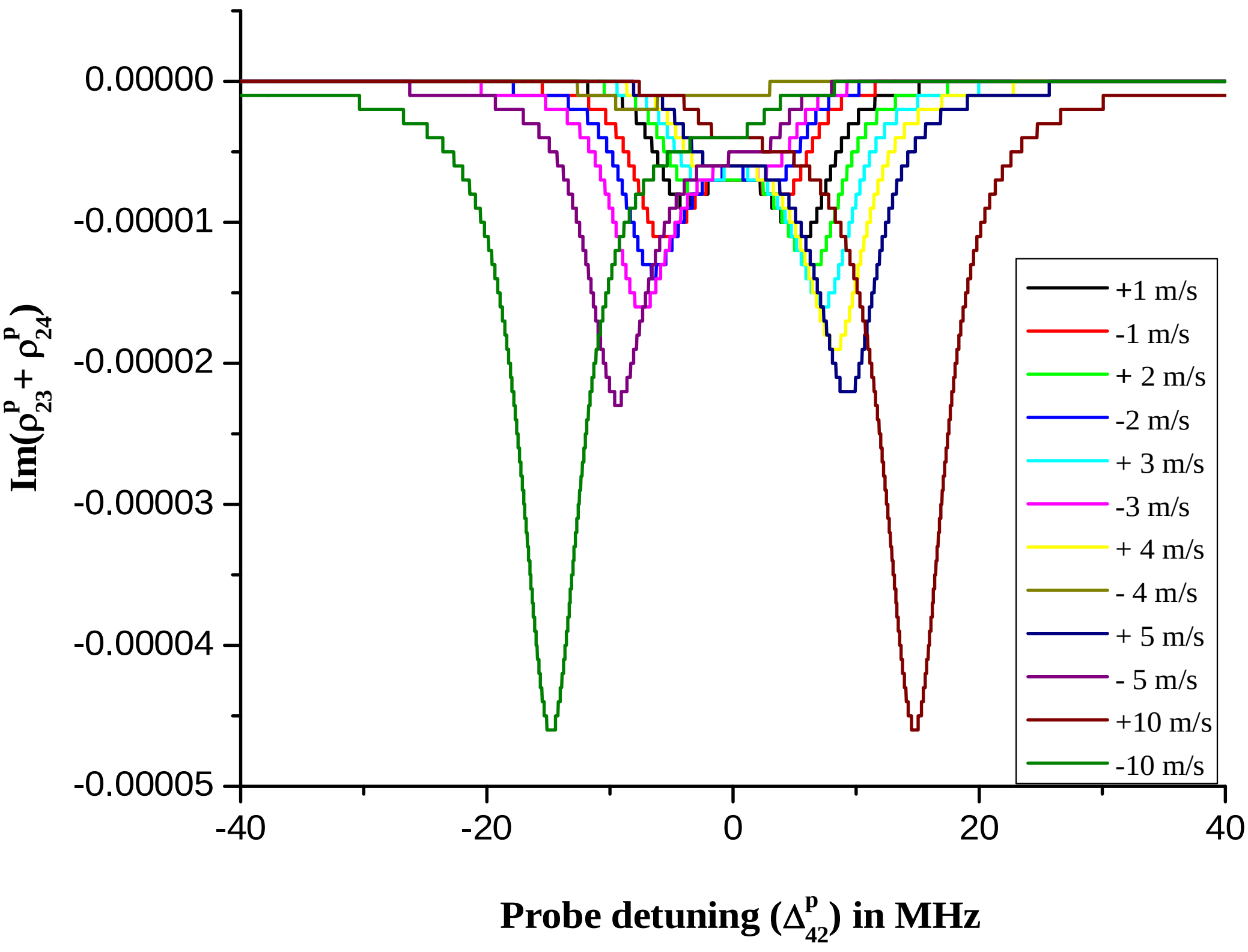

Fig.4: Plot of $Im(\rho_{23}^{p}+ \rho_{24}^{p})$ vs. probe frequency detuning $(\Delta_{42}^{p})$for different velocity groups of atoms (as shown in the figure).

The interaction of the non-zero velocity group of atoms with the probe field causes an increased absorption around $\Delta_{42}^{p}$= 0.The coupling of the probe field to both the upper levels means when it is on resonant with the |2> → |4> transition ($\omega_p = \omega_4 - \omega_2$), it is still sending atoms with velocity 'v' (for which $\omega_4 - \omega_3 = \omega_p(v/c)$) [21] to level |3>. The atoms thus pumped to |3> are allowed to decay spontaneously to both the ground levels |2>and |1>. Under Doppler free situation this extra coupling is not possible. If we does not allow the probe field to couple both the excited states (|3> and |4>, Fig.1) in the theoretical model and allow the probe to couple |2> with |4> only, no such absorption dip in the EIT window at $\Delta_{42}^{p}$ =0 can be found. This has been shown in the fig.5 below for different values of the control Rabi frequency.

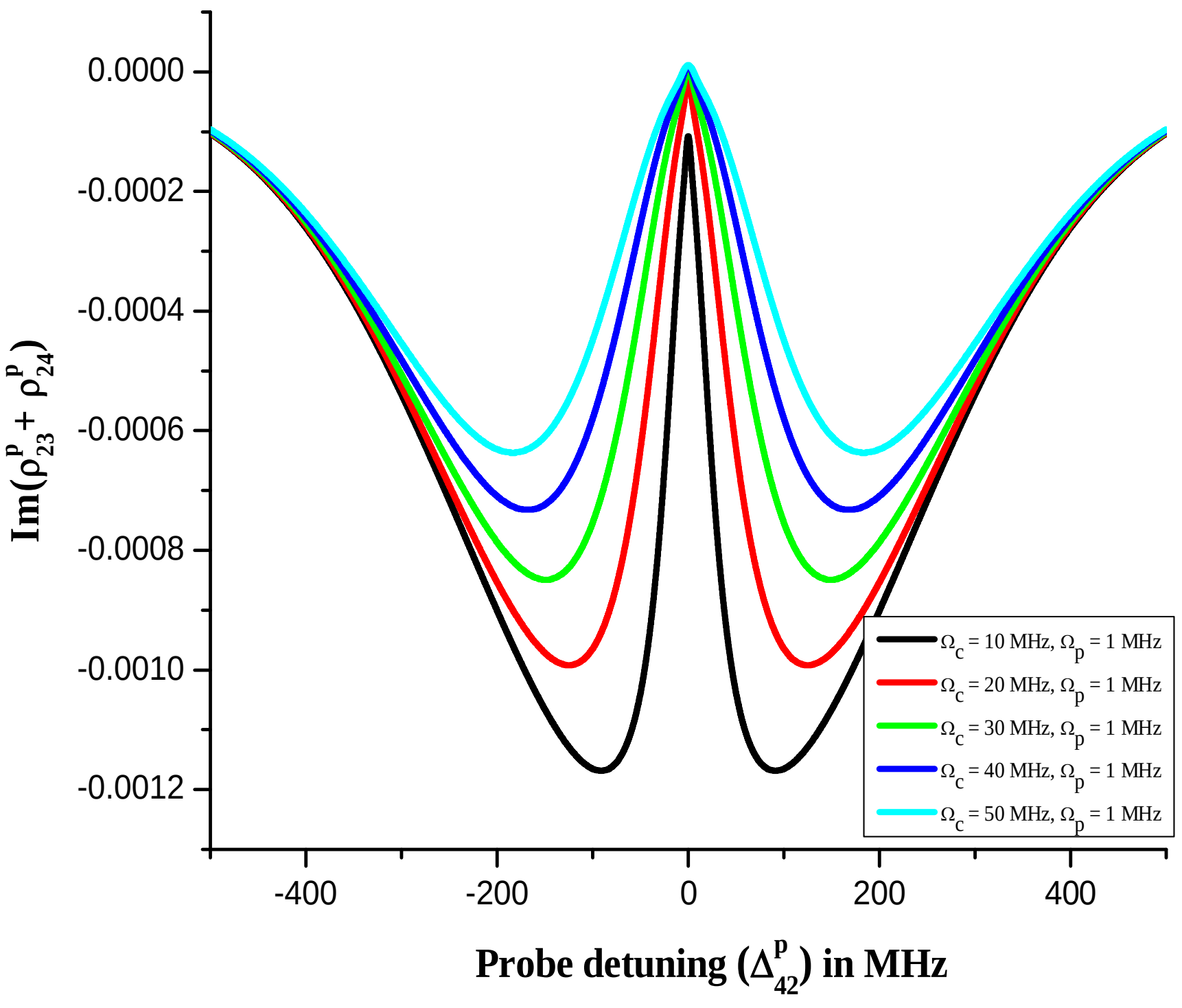

Fig.5: Plot of $Im(\rho_{24}^{p})$ vs. probe frequency detuning $\Delta_{42}^{p}$at different values of control Rabi frequency for 4-level V-type system. Pump couples |2> → |3>, probe couples |2> → |4>.

If we vary the temperature of the ensemble, we observe variation of the width of the field induced absorption (Fig.6). We have also shown the zoomed absorption dip at different temperature in the inset of Fig.6 below.

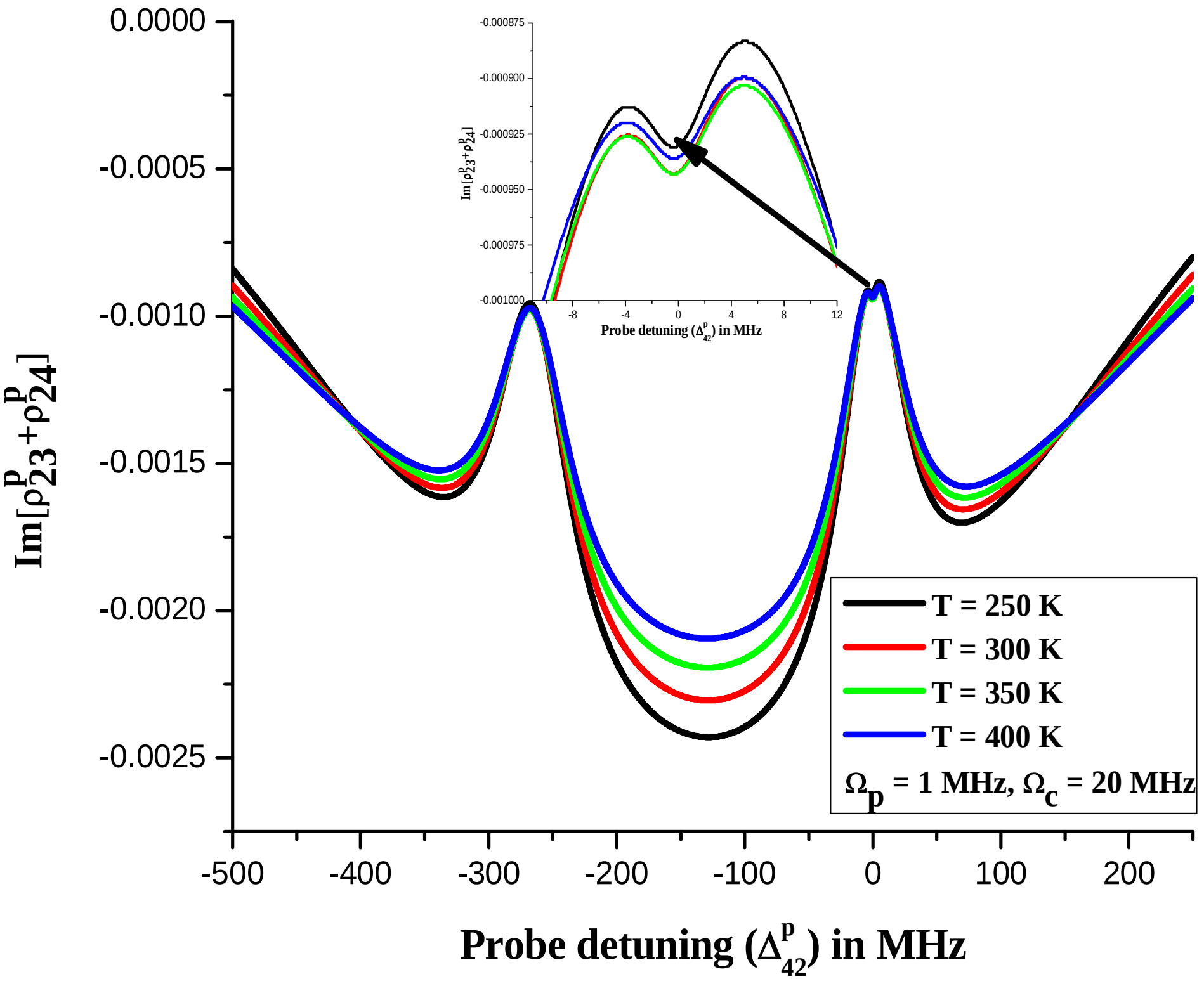

Fig.6: Plot of $Im(\rho^{p}_{23}+\rho^{p}_{24})$ vs. probe frequency detuning $(\Delta^{p}_{42})$ at different temperatures (as shown in the figure). The pump and probe Rabi frequencies are 20 MHz and 1 MHz respectively.

A plot of temperature vs. fitted width of the field induced absorption at fixed values of the control and probe Rabi frequencies ($\Omega_p$ = 1 MHz, $\Omega_c$ = 20 MHz) has been given in the figure below (Fig.7) to establish that thermal broadening is indeed playing a vital role in generating the line shape. The linear variation of the fitted width of the absorption dip with temperature is seen. The effect of the thermal averaging on reducing the line width of the absorption dip is thus confirmed. As usual, no shift of the absorption dip with temperature is found. Increase in the ensemble temperature broadens the Doppler profile.

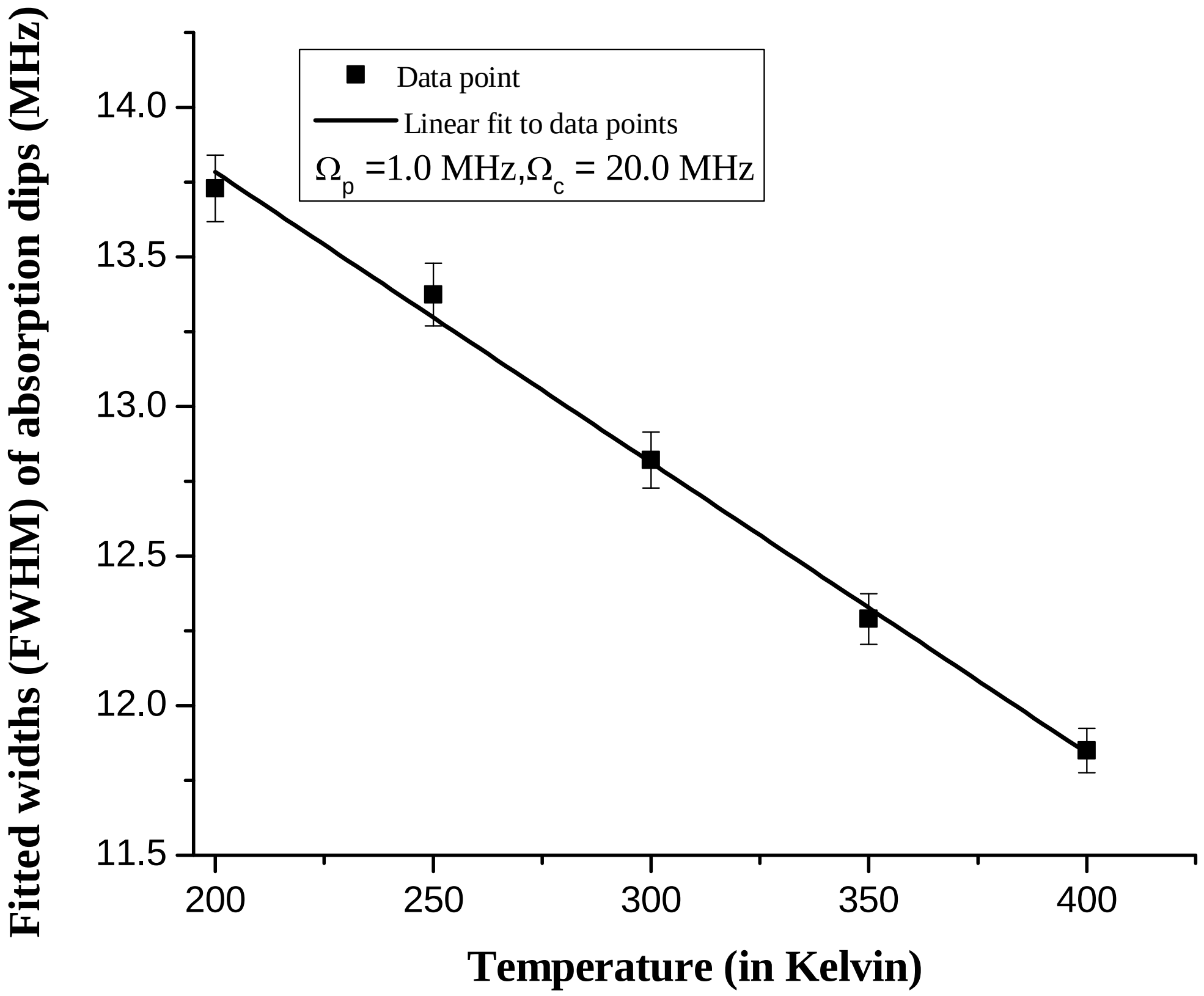

Fig.7: Plot of the fitted widths of the absorption dips vs. temperature. The values of the control and probe Rabi frequencies have been shown in the graph.

A linear shift of the field induced absorption dip with increase in the value of the Rabi frequency of the control field is observed(Fig.8 below) although the fitted width of this dip does not show any linear dependence on the control Rabi frequency (Fig.9 below). The probe Rabi frequency is kept fixed at 1 MHz. The temperature of the system is kept at 300 K. Fig.8 just quantifies the observation made in fig.2 in this regard. The shift of the absorption dip with increase in the control Rabi frequency is attributed to the AC Stark shift [25]. We have not given any analytical expression for the width and shift of the peak since the mathematical expression for the probe response ($\rho_{23}^{p}+\ \rho_{24}^{p}$) is very complicated (Eq.12 and Eq.13). It is very difficult to derive any simple expression for these parameters from the algebraic expressions written in Eq.12 and Eq.13 without any approximation. We have tried to avoid the use of approximation throughout the treatment as far as possible and relied on the graphical representation of the analytically obtained probe response.

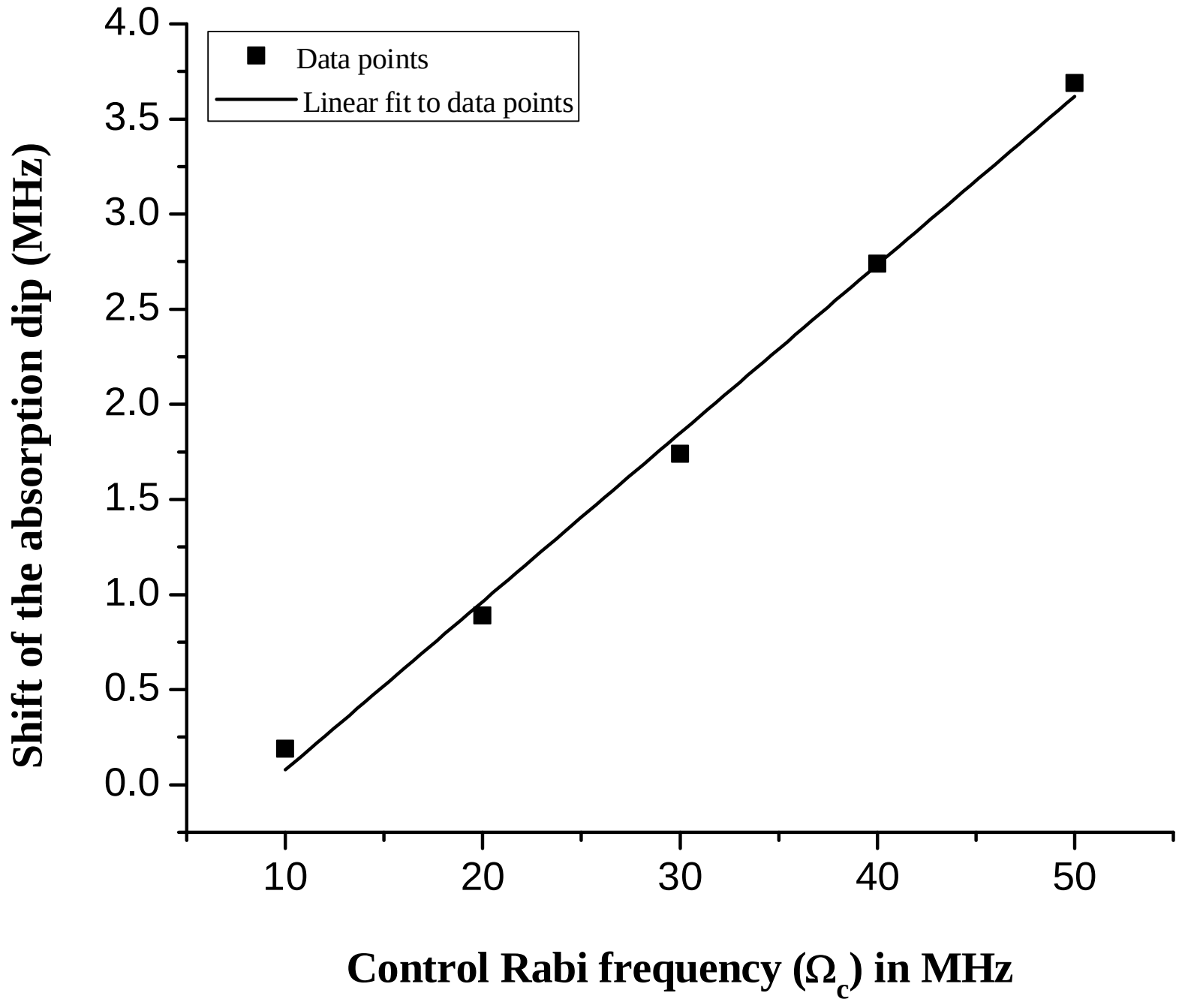

Fig.8: A plot of the shift (in MHz) of the absorption dips vs. control Rabi frequency (in MHz) at T = 300 K. The probe Rabi frequency is held fixed at 1 MHz throughout the simulation.

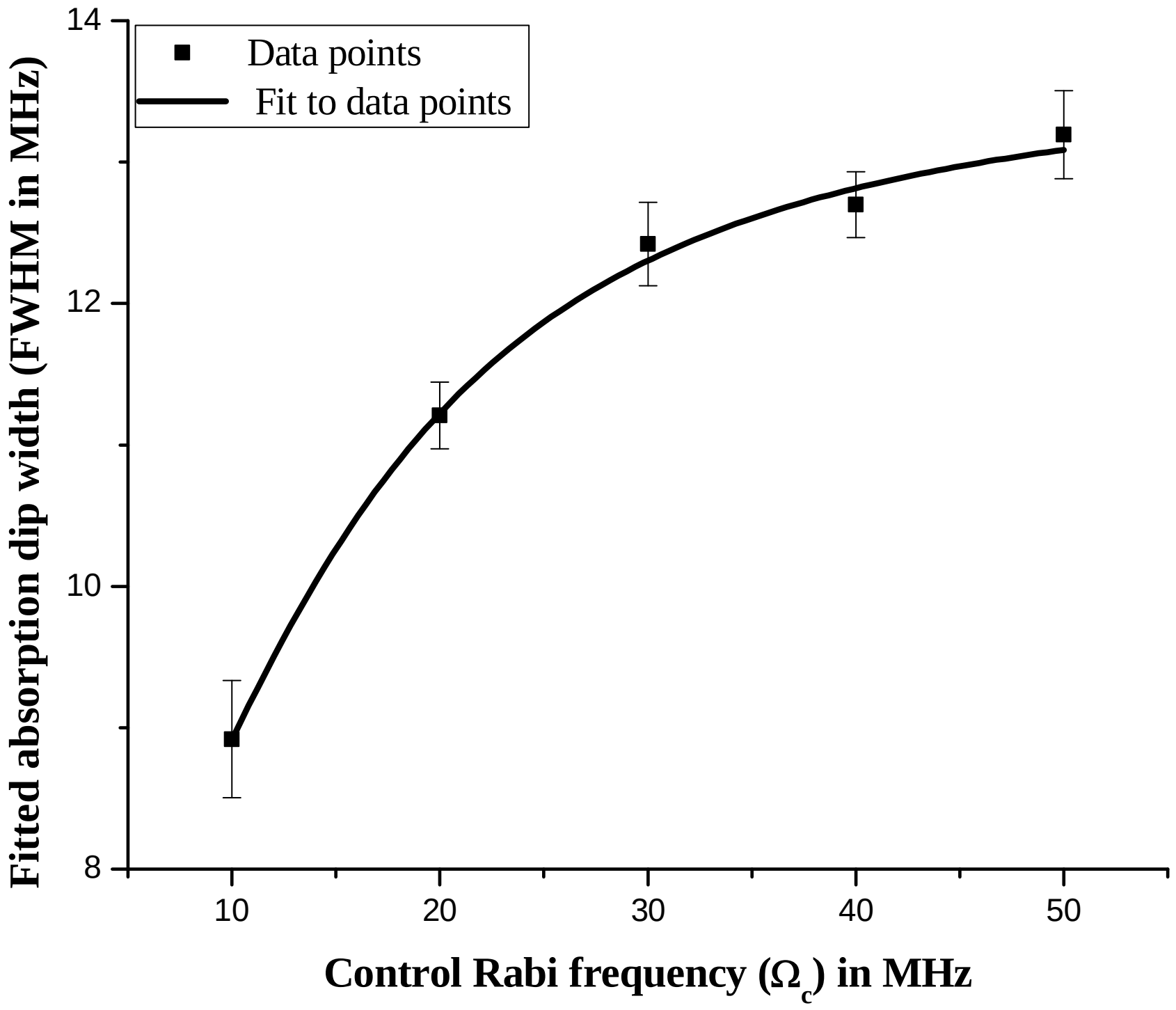

Fig.9: A plot of the fitted widths (FWHM in MHz) of the absorption dips vs. control Rabi frequency (in MHz) at T = 300 K. Probe Rabi frequency is 1 MHz for all the data points.

A plot of the real part of ($\rho_{23}^{p}$+ $\rho_{24}^{p}$) vs. probe detuning ($\Delta_{42}^{p}$) under both the Doppler free and Doppler broadened regimes allows us to get an idea of the dispersive properties of the atomic medium under the two different conditions. These are shown in Fig.10 and Fig.11 below. The steepness of the dispersion at $\Delta_{42}^{p}$ = 0 under the Doppler free condition decreases with the increase in the Rabi frequency of the control field although under Doppler broadened condition we expect modification in the dispersive behaviour of the atomic vapour. This information will be useful to prepare a medium in which a sudden variation in the dispersive property within a very short range is required. The intensity of the control field acting on the energy levels of the atomic vapour system will be one of the main controlling factors of the nature of the dispersion curve.

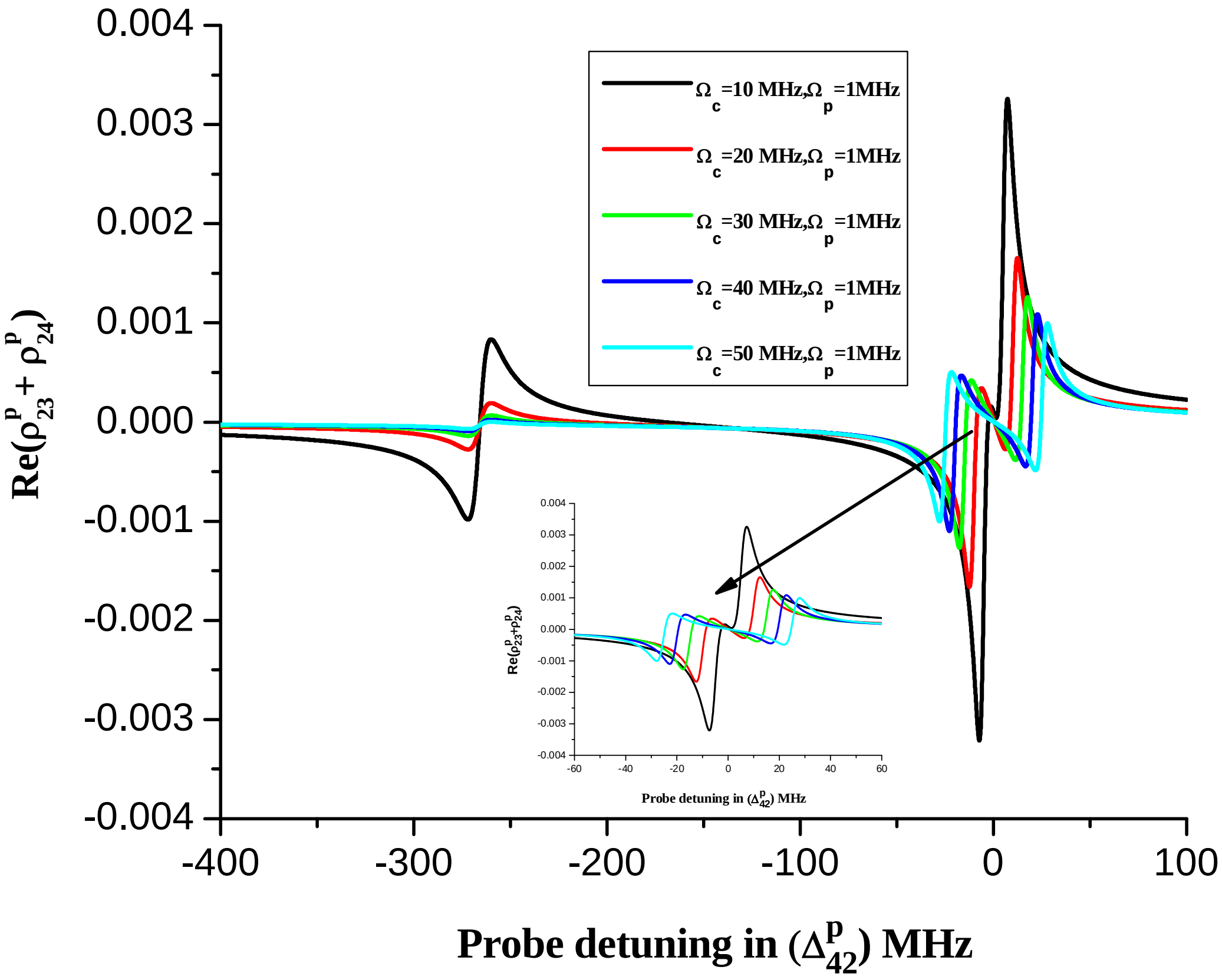

Fig.10: Plot of $Re(\rho_{23}^{p}$+ $\rho_{24}^{p})$ vs. probe detuning (MHz) under Doppler free condition.

It is evident from the above figure that with increase in the Rabi frequency of the control field the probe dispersion will get flatter and flatter. Under Doppler broadened regime the steepness of the dispersion curve however decreases with increase in the Rabi frequency of the control field, a feature just similar to that of Doppler free condition (please compare

Fig.10 and Fig.11). A kink like structure is appearing in the dispersion curve around the zero probe field detuning ( $\Delta_{42}^{p}$= 0, fig. 11 below). This kink becomes prominent at higher intensities of the control field. This has been shown in the inset of Fig.11 below. This kink modulates the probe dispersion around zero probe detuning. This may have useful application in studying the probe pulse propagation through such an atomic vapour system. This will be quantified in a future correspondence.

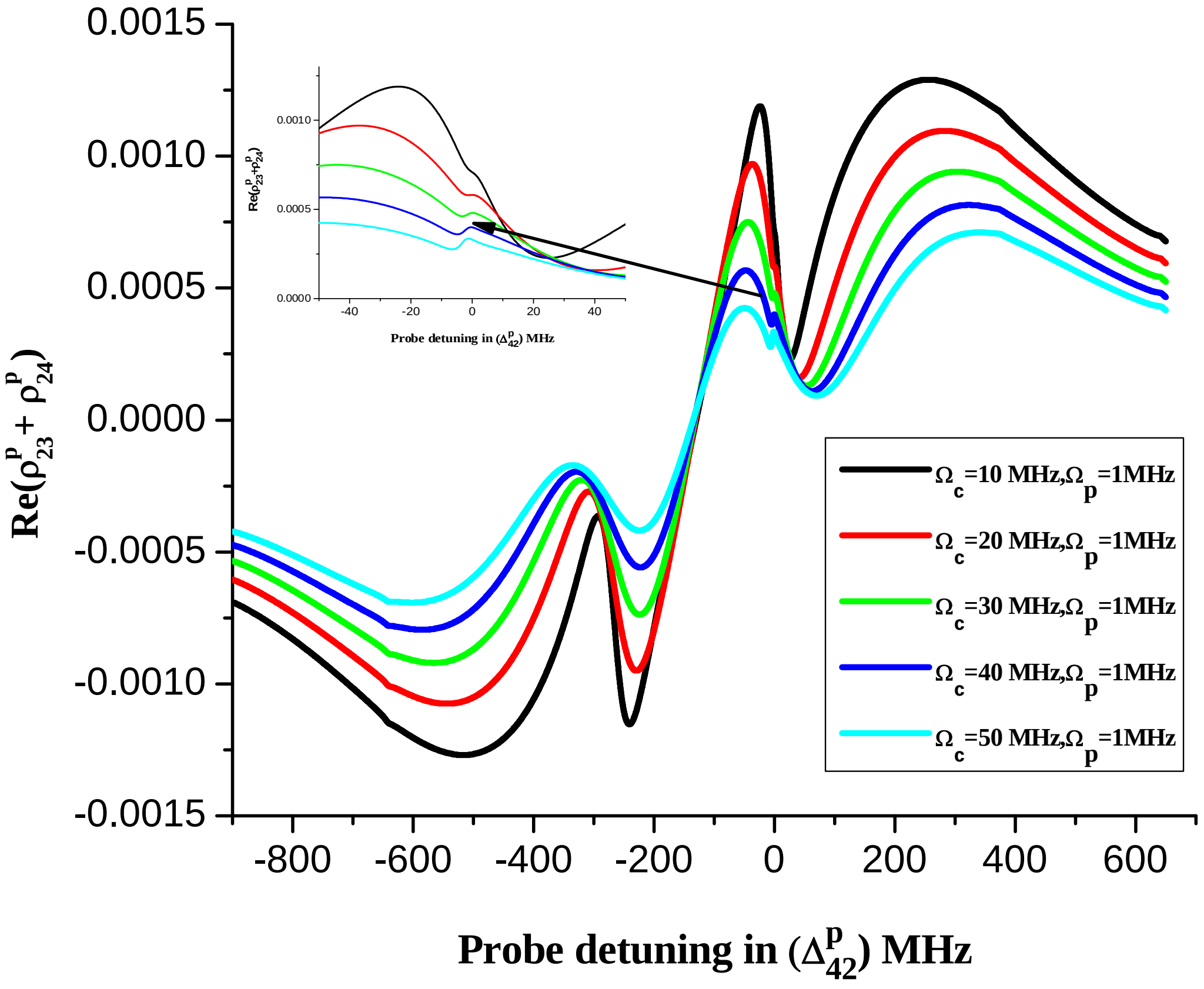

Fig.11: Plot of $Re(\rho_{23}^{p}+ \rho_{24}^{p})$ vs. probe detuning (in MHz) under Doppler broadened conditions.

Till now we have been interested in studying the probe response under both the Doppler broadened and Doppler free conditions by allowing the decay of the population from excited level |3> to both the ground levels |1> and |2> but that from the level |4>has been restricted only to level |2>. If the population from the state |4> is allowed to decay spontaneously to both the ground levels we shall get alteration in the probe response. The appearance of the absorption dip at and around the zero probe field detuning ( $\Delta_{42}^{p}$= 0) in the transparency background as seen in Fig.2 will be absent at lower values of the Rabi frequency of the control field. With increase in the Rabi frequency of the control field we can get back the absorption dip at the same position but with lesser strength as compared to Fig.2. This has been shown in the Fig.12 below.

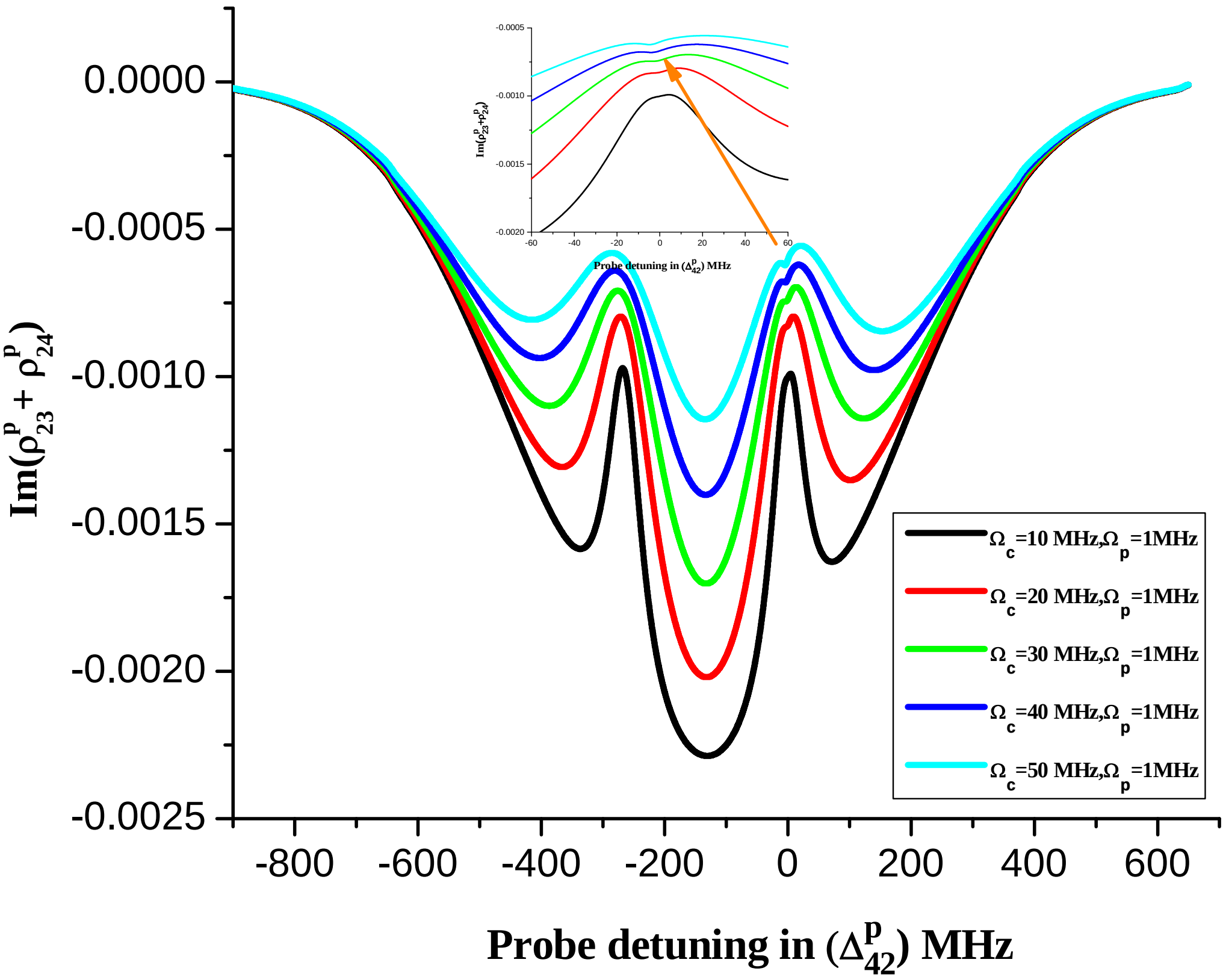

Fig.12: Plot of imaginary part of the ($\rho^p_{23}$+ $\rho^p_{24}$) vs. probe frequency detuning ($\Delta^p_{42}$) under the Doppler broadened regime (T = 300 K) with $\gamma_{41}$ = 6 MHz. The values of the control ($\Omega_c$) and probe ($\Omega_p$) Rabi frequencies have been mentioned in the figure.

In the inset of the above figure we have shown the zoomed probe response around zero probe field detuning. If this is compared (Fig.13 below) with the zoomed part of Fig.2 we can see that the absorption dip starts appearing when the Rabi frequency of the control field is increased to 30 MHz and beyond but the absorption dips are not as prominent as in Fig.2. The introduction of the extra decay channel in the theoretical model enhances the transparency by diminishing the absorption at and around $\Delta^p_{42}$= 0. We can compare the field induced dips formed in the background of the transparency window for both the cases with and without nonzero contribution from $\gamma_{41}$ under Doppler broadened regime.

We have shown in the Fig.13 clearly that for $\gamma_{41}$= 6MHz no absorption dip is created around the zero probe detuning at a control Rabi frequency of 10 MHz whereas the formation of dip, i. e. enhancement of absorption is minimal, if not insignificant, for control Rabi frequencies of 30 MHz and 50 MHz compared to the case with $\gamma_{41}$= 0 under Doppler broadened condition.

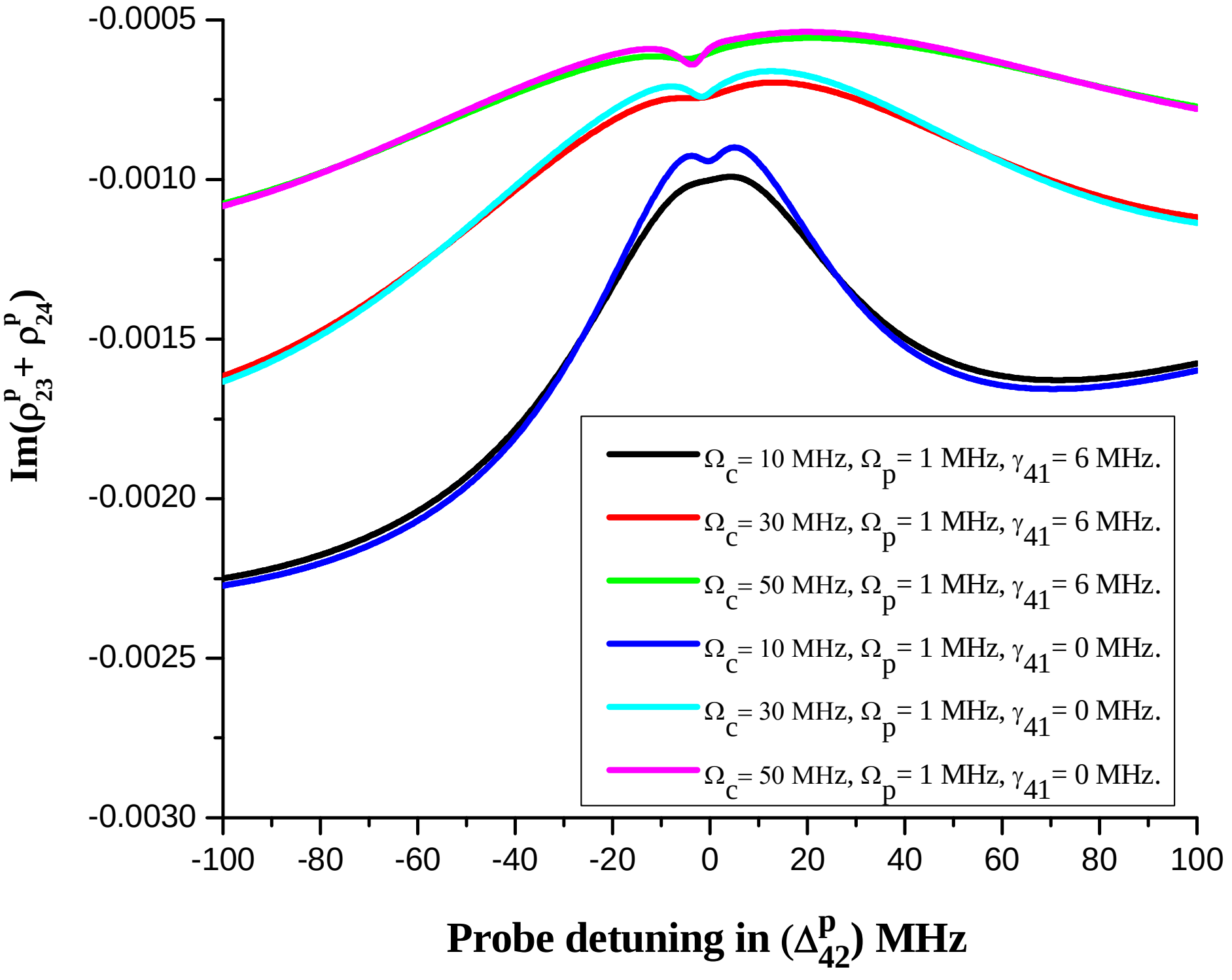

Fig.13: Plot of imaginary part of the $(\rho^p_{23}+\rho^p_{24})$ vs. probe frequency detuning $\Delta^p_{42}$ under the Doppler broadened regime ($T$ = 300 $K$) with $\gamma_{41}$ = 6 MHz as well as $\gamma_{41}$ = 0 MHz. The values of the control ($\Omega_c$) and probe ($\Omega_p$) Rabi frequencies have been mentioned in the figure.

If we investigate the effect of the variation in the population transfer rate between the two ground levels |1> and |2> on the probe response signal, we can have a better understanding of the physical phenomenon responsible in creating the absorption dip. We have already shown that the formation of this absorption dip is related to Doppler broadening (please compare Fig.2 and Fig.3, Fig.3 does not show any enhancement in the probe absorption in the EIT background at zero probe detuning whereas we observe absorption dip in the transparency background at zero probe detuning when Doppler broadening is taken into account in the simulation). If we still keep $\gamma_{31} = \gamma_{32} = \gamma_{42}$ = 6 MHz and $\gamma_{41}$ = 0 and vary the population transfer rates ($\gamma_{12}$ and $\gamma_{21}$) between the ground levels (|1> and |2>) we can clearly see that

the dip appearing at the zero probe detuning is becoming more and more prominent with increase in$\gamma_{12}$. This is shown in the fig.14 below along with the statistics of fits in table-1. For each of the graphs in Fig.14 we used $\gamma_{12} = \gamma_{21}$.

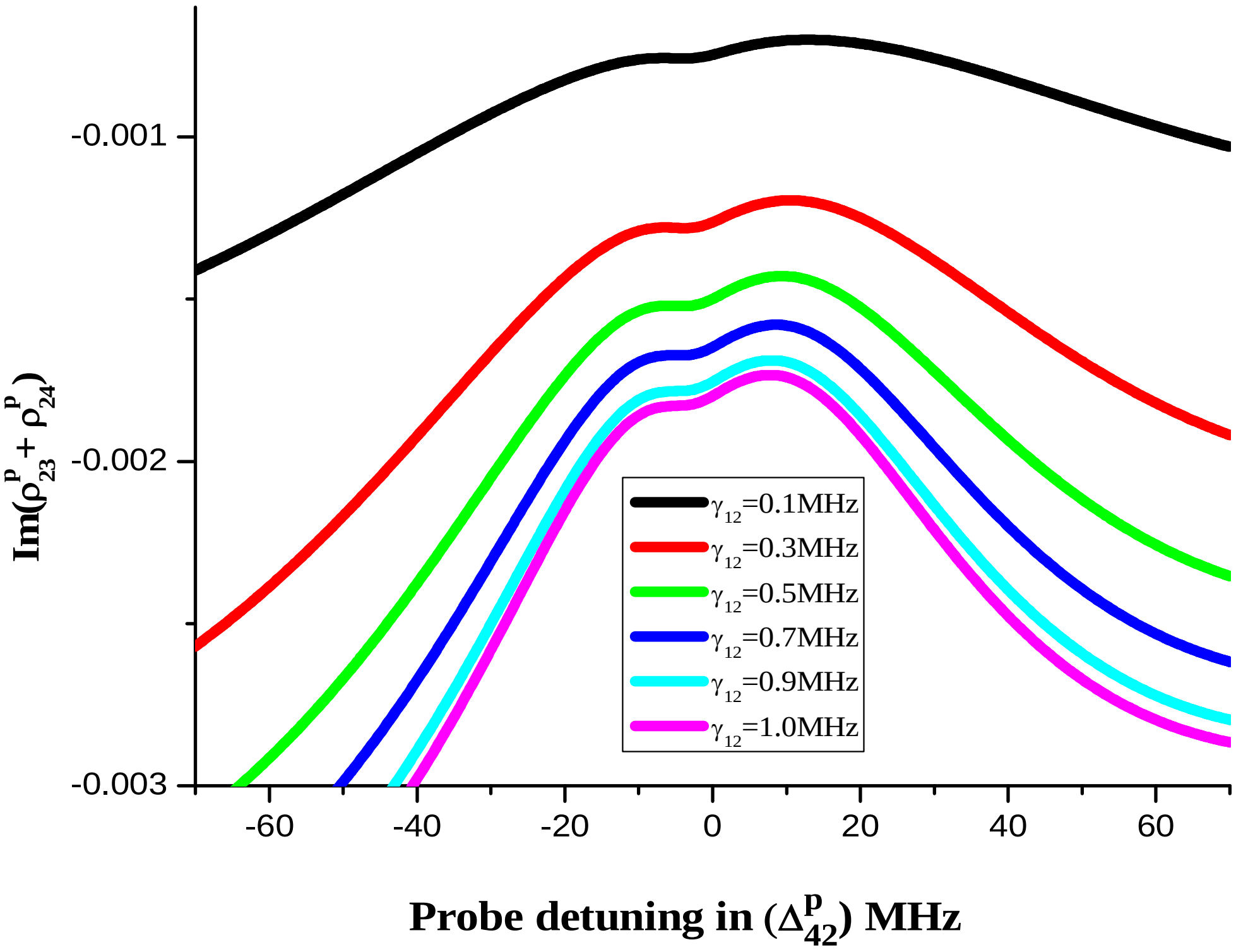

Fig.14: Plot of imaginary part of the ($\rho^p_{23}$+ $\rho^p_{24}$) vs. probe frequency detuning ($\Delta^p_{42}$) under the Doppler broadened regime (T = 300 K) with $\gamma_{41}$ = 0 MHz. The values of the control ($\Omega_c$) and probe ($\Omega_p$) Rabi frequencies have been kept fixed at 1 MHz and 30 MHz respectively. The values of $\gamma_{12}$ (population transfer rate from |1> to |2>) have been mentioned in the figure.

**Table - 1**

| Population Transfer rate $\gamma_{12}$ (from \|1> to \|2>) in MHz | Area under the absorption dip (in arb. unit) | Standard error of fit (Area) | Amplitude of the absorption dip (in arb. unit) |
|---|---|---|---|
| 0.1 | $4.56467\times10^{-4}$ | $6.68900\times10^{-6}$ | $2.63097\times10^{-5}$ |
| 0.3 | $5.92954\times10^{-4}$ | $1.3208\ 0\times10^{-5}$ | $4.06860\times10^{-5}$ |
| 0.5 | $6.87988\times10^{-4}$ | $1.39241\times10^{-5}$ | $4.61169\times10^{-5}$ |
| 0.7 | $7.33530\times10^{-4}$ | $2.40127\times10^{-5}$ | $4.79099\times10^{-5}$ |
| 0.9 | $8.1815\ 0\times10^{-4}$ | $4.0020\ 0\times10^{-5}$ | $5.01683\times10^{-5}$ |
| 1.0 | $8.9389\ 0\times10^{-4}$ | $4.9221\ 0\times10^{-5}$ | $5.2222\ 0\times10^{-5}$ |

The availability of more atoms in the level |2> means these atoms would interact with the coherent fields and thereby reduce the transparency at the zero probe detuning (under Doppler broadened regime) when the population transfer rate from |1> to |2>is increased. In the Fig.15 below we have also shown that the increase of the area under the absorption dip with respect to $\gamma_{12}$is approximately linear. The values of the control and probe Rabi frequencies have been kept fixed at 20 MHz and 1 MHz respectively.

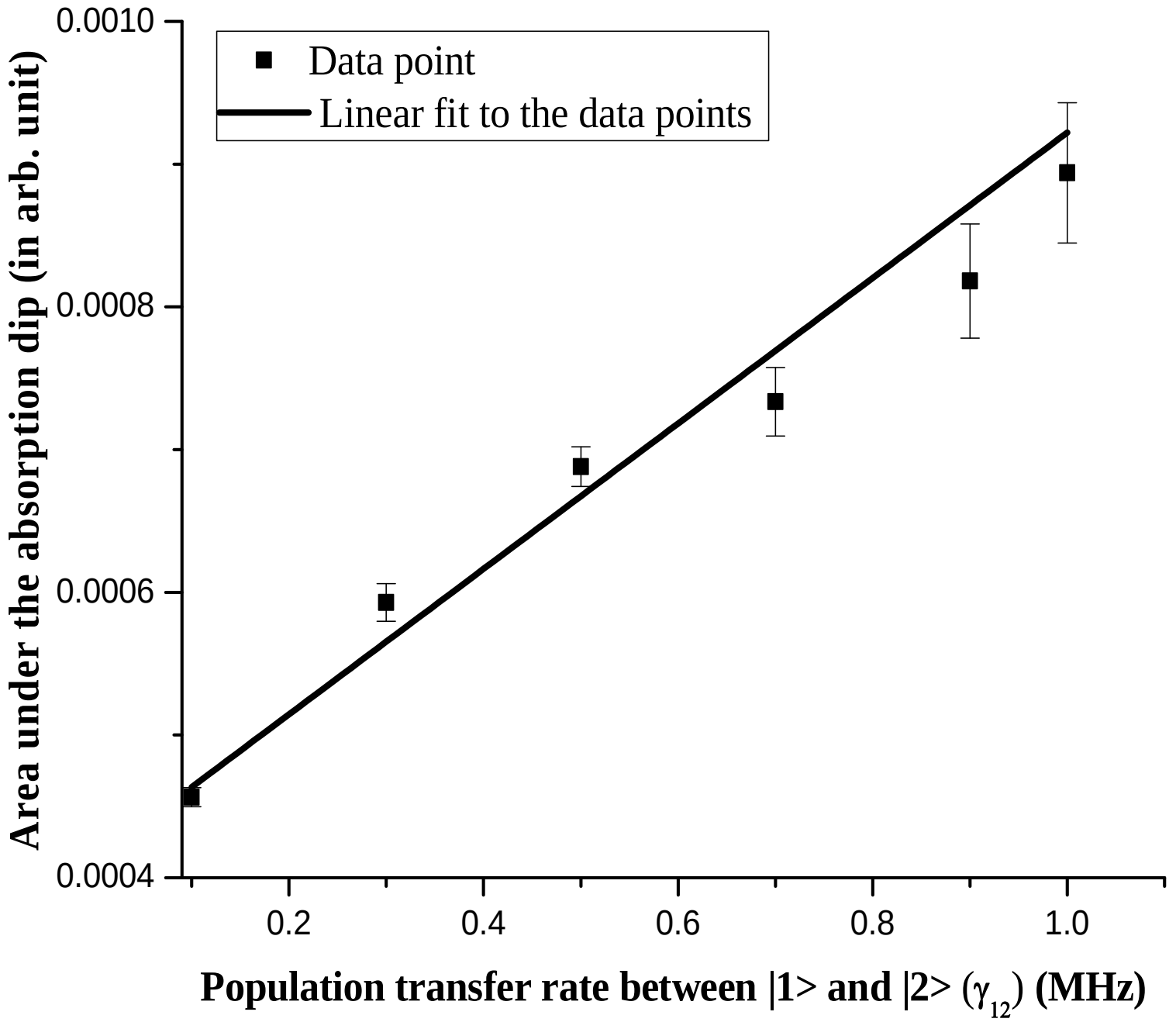

Fig.15: A linear fit to the plot of the area under the absorption dip (appearing around the zero probe detuning) vs. $\gamma_{12}$. The values of the control and probe Rabi frequency used in the simulation are 20 MHz and 1 MHz respectively.

## 4. Conclusion

We have mainly shown here the effect of the addition of an extra ground level in a three-level 'V' type system and the role played by extra decay channels of the excited level population on the probe response. The theoretical model has been developed to incorporate a probe field coupling both the upper levels and a control field locked to a specific transition. The effect of the inclusion and non-inclusion of different possible decay channels on the probe absorption spectra has been shown under various conditions. The effect of Doppler broadening and the formation of a field induced absorption signal on the background of a transparency window on the probe response signal have been shown. The dependence of this field induced absorption on various parameters like temperature, Rabi frequency of the control field etc. has also been investigated in details. In fact, to the best of our knowledge, formation of an absorption dip on the transparency background in a four-level 'V' type system is shown here for the first time through a complete analytical treatment. The approach has been to incorporate the experimental conditions to the closest possible way in the simulation. A scheme is also presented on the basis of the outcome of this theoretical study for producing a dispersive medium where rapid manipulation of the group velocity of the probe pulse propagation can be achieved within a very small range of probe frequency detuning. The model can also be used to study the effect of dephasing and other collision related population transfer phenomena among the ground levels on the probe response. We believe that this simple analytical model will be useful in probing the 'V' type system in further details in the future and exciting new results can be obtained.

**Acknowledgements:** AB acknowledges Department of Science & Technology (DST), New Delhi for granting a research project under the start-up research grant (Young Scientists) (sanction order no. SR/FTP/PS-079/2010, dated: 14/08/2013) and the University Grants Commission (UGC), New Delhi for granting a major research project (F. No. – 43-527/2014(SR) dated: 28/09/2015). DB thanks the University Grants Commission (UGC) for a research project (Sanction order no. F PSW-205/13-14 dated 01/08/2014). KI and AG acknowledge Visva-Bharati for providing research fellowship. The authors thank Prof. Swapan Mandal (Department of Physics, Visva-Bharati) for sharing his experiences and resources.